# Tunnel magnetoresistance in magnetic tunnel junctions with embedded nanoparticles

A.N. Useinov[1-3], N.K Useinov[2], L.-X. Ye[3], T-H. Wu[4] and C.-H. Lai[3*]

[1]Department of Physics, National Tsing Hua University, Hsinchu 30013, Taiwan
[2]Solid State Physics Department, Kazan Federal University, Kazan 420008, Russia
[3]Department of Materials Science and Engineering, National Tsing Hua University, Hsinchu, Taiwan
[4]Graduate School of Materials Science, National Yunlin University of Science and Technology, Douliou, Taiwan

**In this paper, we attempt the theoretical modeling of the magnetic tunnel junctions with embedded magnetic and nonmagnetic nanoparticles (NPs). A few abnormal tunnel magnetoresistance (TMR) effects, observed in related experiments, can be easily simulated within our model: we found, that the suppressed TMR magnitudes and the TMR sign-reversing effect at small voltages are related to the electron momentum states of the NP located inside the insulating layer. All these TMR behaviors can be explained within the tunneling model, where NP is simulated as a quantum well (QW). The coherent (direct) double barrier tunneling is dominating over the single barrier one. The origin of the TMR suppression is the quantized angle transparency for spin polarized electrons being in one of the lowest QW states. The phenomenon was classified as the quantized conductance regime due to restricted geometry.**

## I. Introduction

It is well known that the magnetic tunnel junctions (MTJs) can be used as basic elements for magnetic random access memory [1], as magnetic field sensors for detecting micron-sized particles [2], nanosized devices, and instruments for analyzing the effects of spin pumping [3], and so on. A special type of the MTJs can be fabricated by the sputtering technique, where the layer of nanoparticles (NPs) can be deposited inside the junction's barrier. The embedded NPs modify the properties of the tunnel junctions and usually increase the tunnel magnetoresistance (TMR) amplitude [4-6].

In this paper, we present a theoretical background for the TMR in MTJ with the embedded nonmagnetic and magnetic NPs (npMTJ). The simulation was done within the model of ballistic tunneling through the insulating layer containing NPs. We considered two conduction channels connected in parallel within one MTJ cell: first one is through double barrier subsystem (DBSS), path **I** (Fig.1), and second one is through a single barrier subsystem (SBSS), path **II**. The model allows us to reproduce, for example, the TMR dependencies of the experimental results for npMTJs derived at low and room temperatures having in-plane [4, 5] as well as perpendicular magnetic anisotropy [6]. In the original experimental work [4] the interpretation of these behaviors was determined assuming the existence of the sequential tunneling, cotunneling and nonresonant (and resonant) Kondo-assisted tunneling regimes, depending on conditional thickness of embedded layer, which correlates with $Co_{70}Fe_{30}$ NPs size distribution. In our model, we reproduced the anomalous bias-dependence of the TMR: its suppression and enhancement with the magnetic and non-magnetic NPs, exploring also the temperature factor in range of the one tunneling regime. We found that the electron transport through NPs is similar to coherent tunneling

Corresponding author: Chih-Huang Lai (e-mail: chlai@mx.nthu.edu.tw).
Corresponding author: Arthur Useinov (e-mail: ausein@rambler.ru).

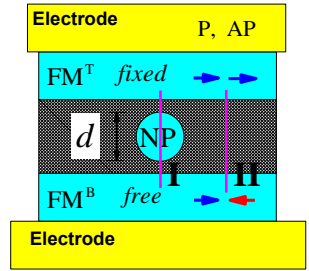

Fig. 1. Schematic view of the symmetric tunneling cell with NP. Arrows show parallel (P) and antiparallel (AP) magnetizations.

in the double barrier MTJ including possible realization of the resonance tunneling cases. The model for these junctions was developed earlier [7, 8].

## II. Theoretical model

In the range of ballistic approximation, the simplified solution for the tunnel current was derived for the systems alike $FM^T$/**Ins**/$FM^B$ & $FM^T$/**Ins**/NP/**Ins**/$FM^B$, where **Ins** is insulator (e.g. MgO, $Al_2O_3$), T and B are the indexes determining top and bottom ferromagnetic layers (FM), respectively. Electrical conductance is proportional to the product of the transmission coefficient (TC) and the cosine $\cos(\theta_{T,s}) \equiv x_s$ of the incidence angle of the electron trajectory $\theta_{T,s}$, Ref. [9], and averaged over solid angle $\Omega$ in spherical coordinates:

$$G_s^{P(AP)}(V) = G_0 \frac{\sigma \cdot (k_{F,s}^T)^2}{2\pi} \langle x_s D_s^{P(AP)} \rangle_\Omega , \qquad (1)$$

where $G_0 = e^2/h$ is the spin-resolved conductance quantum, $e$ and $h$ are the electron charge and Planck's constant, $\sigma$ is cross section area of the tunneling subsystem, $k_{F,s}^T$ is the electron Fermi wavenumber of the top electrode and $s = \uparrow (\downarrow)$ is the spin index; $D_s^{P(AP)}$ is the TC for the single or double barrier system, where TC is a function of the barrier topology: NPs diameter $d$, barriers heights $U_{1,2}$ and widths $L_{1,2}$, applied bias-voltage $V$, values of the wave vectors $k_{F,s}^j$ ($j$= T, NP, B), and effective electron masses $m_j$. Equation (1) applied for positive $V$, while the solution for negative one can be derived using symmetric relations of the system, i.e., the parameters of the majority (maj) and minority (min) electronic states of the contact FM layers have to be reversed $k_{F,s}^{T(B)} \to k_{F,s}^{B(T)}$.

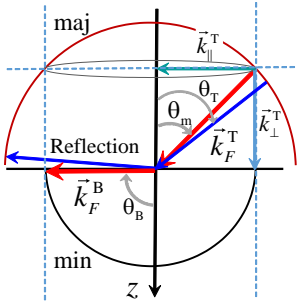

Fig. 2. Angle restriction of the electron trajectory: $\vec{k}_F^T$ shows the transmission ($\theta_T \leq \theta_m$) and reflection ($\theta_T > \theta_m$) cases. The semicircles are maj and min Fermi surfaces of the spin bands.

The TC is defined as a ratio of the transmitted probability density into $FM^B$ to the incident one in $FM^T$. Assuming the probability density for the $FM^T$ equals unity, and there is only a transmitted component in the $FM^B$, $D_s^{P(AP)}$ takes the form:

$$D_s^{P(AP)} = \frac{m_T k_{\perp,s}^B}{m_B k_{\perp,s}^T}\left(\Psi_s^{P(AP)} \cdot \Psi_s^{P(AP)*}\right), \quad (2)$$

where $\Psi_s^{P(AP)}$ is the complex function of the incident electron with spin $s$, (in our calculations $m_{T(B)}$ are equal to the free electron mass $m_e$). This function was derived by solving the system of the wave functions and their boundary conditions (for more details see [7]). Moreover, the TC can be found by transfer matrix technique. The TC depends on the electron trajectory angle $\theta_{T,s}$, according to the conditions for the transverse components (aligned along the normal of the interface, Fig.2) of the electron wave vectors in SBSS:

$$\begin{aligned} k_{\perp,s}^T &= k_{F,s}^T \cos(\theta_{T,s}), \\ k_{\perp,s}^B &= k_{F,s}^B(V)\cos(\theta_{B,s}), \end{aligned} \quad (3)$$

where $k_{F,s}^B(V) = \sqrt{\left(k_{F,s}^B\right)^2 + \left(2m_B e/\hbar^2\right)V}$ is the voltage-dependent absolute value of the Fermi $k$-vectors of the $FM^B$. Calculations for the DBSS were made under the following additional condition:

$$k_{\perp,s}^{NP} = k_s^{NP}(V)\cos(\theta_{NP,s}), \quad (4)$$

where $k_s^{NP}(V) = \sqrt{\left(k_s^{NP}\right)^2 + \left(2m_{NP}e/\hbar^2\right)V/2}$.

The term $\langle(...)\rangle_\Omega$ in eq. (1) by definition is the following:

$$\langle...\rangle_\Omega \equiv \frac{1}{2\pi}\int_0^{\theta_m}\sin(\theta_{T,s})d\theta_{T,s}\int_0^{2\pi}d\varphi(...) = \int_{X_{CR}}^{1.0} x_s D_s^{P(AP)} dx_s.$$

The lower limit $X_{CR}$ is the critical restriction; $\theta_m = \pi/2$ and $X_{CR} = \cos(\theta_m) = 0$ when the electrons tunnel from the minority into the majority conduction band, and when electron tunnel from the majority into the minority one (Fig. 2) then for SBSS: $X_{CR} = \sqrt{1-\left(k_{F,maj}^T/k_{F,min}^B\right)^2}$, while for the DBSS: $\theta_m$ is the smallest value between $\theta_1$ and $\theta_2$ angles, where

$$\theta_1 = \arccos\left(\sqrt{\left|1-\left(k_{F,maj}^T/k_{min}^{NP}\right)^2\right|}\right)$$

$$\theta_2 = \arccos\left(\sqrt{\left|1-\left(k_{F,maj}^T/k_{F,min}^B\right)^2\right|}\right).$$

These conditions are the result of the conservation of the longitudinal $k$-components, Fig. 2:

$$k_{\parallel,s}^T = k_{F,s}^T \sin(\theta_{T,s}) = k_{F,s}^B(V)\sin(\theta_{B,s}) \quad (5)$$

for the SBSS, and

$$k_{\parallel,s}^T = k_{F,s}^T \sin(\theta_{T,s}) = k_s^{NP}(V)\sin(\theta_{NP,s}) = k_{F,s}^B(V)\sin(\theta_{B,s}) \quad (6)$$

for the DBSS.

The parallel circuit connection of the tunneling unit cells was employed, where each cell contains one NP per unit cell's area, Q, in our calculations $Q = 20\,\text{nm}^2$, while tunnel junction itself has total surface area $S$ and consists of $N$ cells ($N = S/Q$). The total conductance of the junction is $G = N \times (G_1 + G_2)$, where $G_1 = G_{1,\uparrow} + G_{1,\downarrow}$ is dominant conductance through the tunneling cell with the NP (path I, see Fig.1), $\sigma = \pi d^2/4$; $G_2 = G_{2,\uparrow} + G_{2,\downarrow}$ is conductance of the direct tunneling through the single barrier (path II), $\sigma = (Q - \pi d^2/4)$. Finally $\text{TMR} = (G^P - G^{AP})/G^{AP} \times 100\%$.

### III. CALCULATIONS AND RESULTS

#### A. TMR-V Anomalies at Zero-Voltage Region

Dramatically modified TMR voltage behaviors are shown in Fig. 3, that is a result of the variation of only one parameter: $k^{NP}$ wavenumber value. Figs. 3(a) and (b) shows the TMR for npMTJs with $d = 1.2$ nm and $d = 2.6$ nm, respectively. The initial parameters are fixed for both sides of the interface: $k_{F,\uparrow}^{T(B)} = 1.09\,\text{Å}^{-1}$, $k_{F,\downarrow}^{T(B)} = 0.421\,\text{Å}^{-1}$, $L_{1,2} = 1.0$ nm with barrier heights $U_{1,2} = 1.2$ eV over Fermi level, effective masses $0.4\,m_e$ for MgO barriers and $m_{NP} = 0.8\,m_e$ for the NP. For simplicity, the TMR amplitude at zero voltage ($V \sim 10^{-4}$) was determined as $\text{TMR}_0$. Note that only curves 1 and 3 in

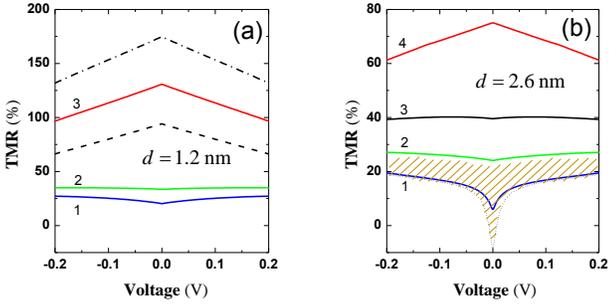
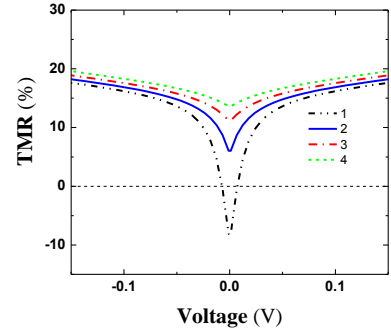

Fig. 3. (a) TMR versus voltage, where curves 1-3 correspond to $k^{NP} = 0.262$ $(n=1)$ $0.3$, and $0.524$ Å$^{-1}$ $(n=2)$, respectively. Black dashed-dotted line is the result of magnetic NP with $k_\uparrow^{NP} = 0.53$, $k_\downarrow^{NP} = 0.518$ Å$^{-1}$, for the black dashed curve: $k_\uparrow^{NP} = 0.518$ $k_\downarrow^{NP} = 0.53$ Å$^{-1}$; (b) Curves 1-4 correspond to $k^{NP} = 0.121$ $(n=1)$, $0.22$, $0.362$ $(n=3)$, and $0.42$ Å$^{-1}$, respectively.

Fig. 4. TMR-$V$ curves close to the QW state $n = 1$. Curves 1-4 correspond to $k^{NP} = 0.115$, $0.121$, $0.127$ and $0.133$ Å$^{-1}$, respectively.

Fig. 3(a) and (b) exactly satisfy the quantum well (QW) solution with $k^{NP} = n\pi/d$ as initial state, where $d$ is the QW width, which equals to the diameter of the NP. The dashed and dot-dashed curves in Fig. 3(a) correspond to magnetic NP in the npMTJ, while all other data are the cases for nonmagnetic NPs. Actually, the NP does not have the ideal geometrical shape and does not clearly satisfy by the QW approach. Thus the relation $k^{NP} = n\pi/d$ can be more complicated, and the $k^{NP}$ can be close to these values. Interestingly, we found that one of the TMR anomalies such as TMR suppression at zero voltage is related to the lowest quantum state $n = 1$. This effect takes place due to the boundary selection rules eq. (6) leading to the quantized conductance. The classical TMR behavior, which corresponds to the enhanced $TMR_0$, takes place for $n \geq 2$.

In the range of our consideration of the magnetic NPs, the approach of the magnetic moment coupling between NP and FM$^B$ is applied. Two opposite types of coupling correspond to coaligned and antialigned magnetic moment orientations of the NP in relation to magnetization alignment of the free (soft) magnetic layer FM$^B$ (including P and AP cases). In Fig. 3(a) the TMR-$V$ solutions are shown for these two cases: 1) dash-doted curve for coaligned and 2) dashed one for antialigned coupling. The $k^{NP}$ are close to the QW state with $n = 2$ (for $d$=1.2 nm), but now $k^{NP}$ is spin-resolved value. The depicted large difference of the TMR amplitudes for these two cases is possible by the cause of the strong dependence between the dominated spin-up channel transparency and $k_\uparrow^{NP}$ value. From our point of view, the experimental data, presented in Ye's work (see [6, Fig. 5]), confirm the existence of the coaligned and antialigned coupling of the magnetic NPs and shows the TMR amplitudes in 111% and 85 %, demonstrating the related resistance switching via external field.

Thus, the standard quantum mechanical solution for an electron tunneling in npMTJs reproduces anomalous voltage behaviors, where the state $n = 1$ can be determined as the quantized conductance regime (see section III-C). For the real systems, however, it is important to consider the size dispersion of NPs, and in addition, consider possible resonant oscillations of the TMR values with $d$, [7]. In present work, the model is limited to an approach of the averaged NP by size per tunneling cell.

Our model explains the experimental observations in [4] and [5] of the $TMR_0$ suppression, which depends on the NP size and temperature factor. However, we cannot observe the peak-like enhancement of the $TMR_0$, which was founded for certain thickness of the middle layer on experiments (see [4, Fig. 2]), where the deposition of the middle layer is responsible for the NPs fraction formation. We assume that the peak-like TMR voltage dependence can be a result of the resonance tunneling and the beginning of the conductance quantized regime [9].

B. *Temperature factor*

The temperature is an important factor, which may determine the TMR-$V$ behavior, for example, the suppressed $TMR_0$ behavior is possible only at low temperatures around a few kelvins. Note, the temperature-induced band broadening at room temperature in $k_B T_0 \approx 0.026$ eV corresponds to $k^{NP}$ margin of about $\pm 0.037$ Å$^{-1}$ ($m_{NP} = 0.8 m_e$), that is comparable with small $k^{NP} \simeq 0.15$ Å$^{-1}$ itself. The room temperature factor for the bulk values of $k^{NP} \simeq 1.0$ Å$^{-1}$ gives much smaller relative impact. The dashed area in Fig. 3(b) shows the TMR curve margin affected by room temperature, but here we used an approach, where temperature does not affect the $k$-values in the FM layers itself. Noticeably, assuming that the conducting states with $k^{NP} < 0.121$ Å$^{-1}$ are allowed for the NPs in tunnel junction, then negative values of $TMR_0$ can be achieved, that perfectly correlates with experiments [4]. In Fig. 4 we show the TMR-$V$ behaviors with small variation of $k^{NP}$ value around the QW state $n = 1$ (curve 2). The curve 1 shows negative TMR at small voltage range, which take place at $k^{NP} = 0.115$ Å$^{-1}$. The temperature-induced band broadening at low temperature $T = 2.5$ K for the curve 2 gives the narrow $k^{NP}$ margin $\pm 3.4 \times 10^{-3}$ Å$^{-1}$ ($k_B T \approx 2 \times 10^{-4}$ eV), that keeps the TMR amplitudes approximately between curve 1 and curve 3.

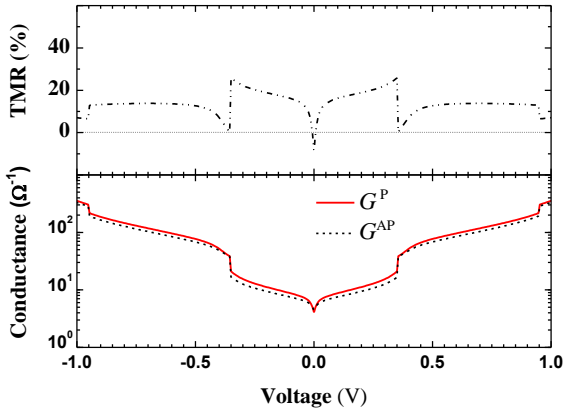

Fig. 5. Presented TMR corresponds to the curve 1, shown in Fig. 4, within extended voltage range. The step-like conductance for this case is shown for the npMTJ with $S = 700 \times 700$ μm, $d = 2.6$ nm.

Furthermore, we found that there is the critical $k_{cr}^{NP}$ value, and when $k^{NP} < k_{cr}^{NP}$ the NP becomes transparent for the conducting electrons, electrons wavelength $\lambda = 2\pi / k^{NP}$ is too large to feel the NP or any QW state. The curve 1 in Fig. 4 shows nonresonant $TMR_0 = -8\%$ for $d = 2.6$ nm and $k_{cr}^{NP} \simeq 0.115$ Å$^{-1}$. The same effect is allowed for the NPs with smaller diameters, where amplitude of the negative $TMR_0$ is smaller, for example, $TMR_0 \simeq -3.8\%$ for $d = 1.2$ nm, and $k_{cr}^{NP} \simeq 0.226$ Å$^{-1}$.

### C. Quantum Conductance Regime

The conductance dependencies for P and AP cases depicted in Fig. 5 show the reason for the $TMR_0$ suppression. The origin of the effect is step-like $G^P$ and $G^{AP}$ conductance behavior, where both the conductance steps (thresholds) are located directly in the region of zero voltage. The conductance step is related to threshold values of the applied voltage. The voltage value directly correlates with the angle restriction of the quantized electron trajectory and reflects the strict conditions of the conservation of the longitudinal components of the $k$-vectors. By the reason of quantized electron trajectory, the effect was classified as quantized conductance regime. The quantum mechanical solution for the $\Psi_s^{P(AP)}$ is stationary by time, coherent in space and corresponds to the standing wave approach. This stationary solution for TC gives the rapid tunnel transparency growth due to opening additional section of the permitted solid angle $\Omega$. It means that there are conditions, since $k^{NP}(V)$ increases with voltage, for which the degree of electrons angle blockade takes down (removes) and it shifts the conductance up for both magnetic configurations. As a result of this rapid growth, $G^P$ becomes comparable with $G^{AP}$ (see Fig. 5).

### IV. CONCLUSION

Thus, the developed model successfully explains some of the experimental TMR anomalies at the zero-voltage region. The simulations also predict the TMR-$V$ amplitude switching due to the existence of the two opposite NPs magnetic moment states: 1) coaligned and 2) antialigned coupling. Moreover, we found that that the quantized conductance is a reason of the $TMR_0$ suppression. Theoretical approach is promising as a background, which allows generalize the model and describe the TMR behaviors in npMTJs more precisely taking into account the dispersion of the NP by size. In this paper, the results were derived only within the quantum mechanical solution and quantum-ballistic electron transport approach. Consecutive tunneling, Kondo-assisted states, Coulomb blockade, and capacitance effects were not considered.

### ACKNOWLEDGMENT

The work was supported by MOST (grant No. 103-2112-M-007-011-MY3) and RFBR (grant No. 14-02-00348-a). A. Useinov acknowledges partial support by the Program of Competitive Growth of Kazan Federal University.